\documentclass[prl,reprint,amsmath,amssymb,showpacs,superscriptaddress ,twocolumn]{revtex4-1}

\usepackage[dvipdfmx]{graphicx}
\usepackage{bm}
\usepackage{mathrsfs}
\usepackage{color}

\DeclareMathOperator{\trace}{Tr}

\DeclareMathOperator{\diag}{diag}

\DeclareMathOperator{\sgn}{sgn}
\DeclareMathOperator{\pf}{Pf}

\begin{document}
\title{Symmetry-protected line nodes and Majorana flat bands in nodal crystalline superconductors}
\author {Shingo Kobayashi}
\affiliation{Institute for Advanced Research, Nagoya University, Nagoya 464-8601, Japan}
\affiliation{Department of Applied Physics, Nagoya University, Nagoya 464-8603, Japan}
\author{Shuntaro Sumita}
\affiliation{Department of Physics, Kyoto University, Kyoto 606-8502, Japan}
\author{Youichi Yanase}
\affiliation{Department of Physics, Kyoto University, Kyoto 606-8502, Japan}
\author{Masatoshi Sato}
\affiliation{Yukawa Institute for Theoretical Physics, Kyoto University, Kyoto 606-8502, Japan}

\date{\today}

\begin{abstract} 
Line nodes in the superconducting gap are known to be a source of Majorana flat bands (MFBs) in time-reversal-invariant superconductors (SCs). Here, we extend this relation to all symmetry-protected line nodes where an additional constraint arising from a symmetry of the crystal destabilizes or hides the existence of MFBs. By establishing a one-to-one correspondence between group theoretical and topological classifications, we are able to classify the possible line-node-induced MFBs, including cases with (magnetic) non-symmorphic space groups. Our theoretical analysis reveals a new type of MFB, i.e., MFBs in antiferromagnetic SCs.
\end{abstract}
\pacs{}
\maketitle

Over the last few years, the study of the nodal gap structures of superconductors (SCs) has experienced renewed interest due to the fact that they can reveal novel kinds of topological objects ~\cite{Volovik03,Horava05,Sato06,Beri10,Zhao13,Shiozaki14,SAYang14,Chiu14,Kobayashi14,Morimoto14,Goswami15,Kobayashi16,Agterberg17,Sato16,Mizushima16,Zhao16,Chiu16,Sato17,Bzdusek17}. In this context, the structures of the superconducting gap are related to the topology of quasiparticles in superconducting states and can ensure the existence of zero-energy surface Andreev bound states~\cite{Hu94,Matsumoto95,Tanaka02,Tanaka12}. Moreover, a line node in time-reversal (TR) invariant SCs is protected by a one-dimensional (1D) topological number and induces a Majorana flat band (MFB) on its surface~\cite{Sato11,Schnyder11,Matsuura13,Schnyder15}. These MFBs exhibit a zero-bias conductance peak that can be revealed through tunneling measurements in TR invariant SCs such as high-$T_{\rm c}$ cuprate SCs~\cite{Tanaka95,Alff97,Lofwander01,Wei98,Iguchi2000,Biswas02,Chesca08} and non-centrosymmetric SCs~\cite{Brydon11,Yada11,Schnyder12}, and thus provide conclusive evidence for the existence of bulk topological line nodes.

In materials with strong spin-orbit interactions (SOIs), e.g., heavy fermion SCs, the formation of Cooper pairs is constrained by the underlying crystal structure because a group operation is followed by the spin operation~\cite{Anderson84,Sigrist91}. When crystal symmetry forbids an irreducible representation (IR) of Cooper pairs in a highly symmetric subspace of the Brillouin zone (BZ), a symmetry-protected node arises~\cite{Sigrist91,Blount85,Volovik84,Volovik85,Ueda85,Yarzhemsky92,Yarzhemsky98}.
Recently, there has been much effort devoted to exploring such symmetry-protected line nodes in SCs with (magnetic) non-symmorphic space group symmetry~\cite{Micklitz09,Mickliz17PRB,Nomoto17,Mickliz17PRL}; however, the corresponding physical phenomena remains unclear. Nevertheless, some of symmetry-protected line nodes are known to be simultaneously protected by a topological number~\cite{Kobayashi16,Mickliz17PRL}, which implies the existence of MFBs.

In this letter, we unify the topology of symmetry-protected line nodes and MFBs. By taking the general symmetry constraints affecting line nodes into account, we show that symmetry-protected line nodes host two different kinds of topological numbers: a zero-dimensional (0D) topological number that describes the topology of the symmetry-protected line nodes, and a 1D topological number that reflects the MFBs. The two topological numbers are intrinsically related to each other and the relationship between them fills the gap between the symmetry-protected line nodes and MFBs. 

Our topological argument allows us to categorize line nodal SCs into three different classes with respect to the MFBs: (i) odd-parity SCs with TR or a magnetic translation symmetry, (ii) even-parity SCs with TR symmetry, and (iii) even-parity SCs with a magnetic translation symmetry. The three classes are directly linked to the symmetry-protected line nodes, as shown in Table~\ref{tab:group} and \ref{tab:topo}. Furthermore, we show that there is no MFB in Class (i) SCs , and that the magnetic translation is sensitive to the surface orientation, which provides an additional constraint on the MFBs in Class (iii) SCs. As a result, each case is distinguishable through surface sensitive measurements. 
Finally, we demonstrate the existence of a MFB in Class (iii) SCs based on a minimal model describing UPd$_2$Al$_{3}$, which hosts nodal loops protected by a magnetic translation symmetry on the Brillouin zone face (ZF). Interestingly, the MFB in this case only arises if the crystal symmetry protecting the nodal loops is broken, reflecting the additional constraint mentioned above.

{\it Irreducible representations of Cooper pairs}--- First, we revisit the group theoretical results. In materials with strong SOIs, electron states respect the symmetry of the crystal structure and are thus characterized by its IRs. Therefore, the possible formation of Cooper pairs is determined by the IRs of the electron states. For clarity, let $\gamma_{\bm{k}}(m)$ be an IR of the symmetry group $m \in G_{\bm{k}}$ for an electron with momentum $\bm{k}$, where $G_{\bm{k}}$ is the little group of $\bm{k}$. When electrons at $\bm{k}$ and $-\bm{k}$ form a Cooper pair, it can be represented by $P_{\bm{k}} (m) \equiv \gamma_{d \bm{k}} (m) \otimes \gamma_{\bm{k}} (m) - \gamma_{\bm{k}} (m) \otimes \gamma_{d \bm{k}} (m)$, where $d$ is a symmetry operator satisfying $d\bm{k}=-\bm{k}$. Therefore, $P_{\bm{k}}$ is an induced representation in $G_{\bm{k}} + d G_{\bm{k}}$, and such anti-symmetrized IRs can be systematically calculated using the Mackey--Bradley theorem~\cite{Bradley70,Bradley72},
\begin{subequations}
\label{eq:MB} \begin{align}
 &\chi [ P_{\bm{k}} (m)] = \chi [\gamma_{\bm{k}} (m)]  \chi [\gamma_{\bm{k}} (d^{-1} m d) ], \label{eq:MB1} \\
 &\chi[ P_{\bm{k}} (dm)] = -\chi [\gamma_{\bm{k}} (d m d m) ], \label{eq:MB2}
\end{align} 
\end{subequations} 
where $\chi$ is a character of the representation. In the following, we apply Eq.~(\ref{eq:MB}) to the case of SCs with line nodes. In the case of a three-dimensional (3D) BZ, a line node may occur along the intersection of a highly symmetric plane and the Fermi surface, so we must also take the mirror-reflection (MR) operation $\sigma_h$ into account as a crystal symmetry. If a Cooper pair lies on such a mirror plane, $d$ then corresponds to either spatial inversion $I$, a two-fold rotation $C_2$, or TR $\theta$. However, in order to take the non-symmorphic group symmetry into account, we must generalize the group operations above so that $\hat{\mathcal{P}} = \{I | 0\}$, $\hat{\mathcal{T}} = \{\theta | \bm{t}_{\theta}\}$, $\hat{\mathcal{M}} = \{\sigma_h | \bm{t}_{\sigma}\}$, and $\hat{\mathcal{C}}_2 = \hat{\mathcal{P}}\hat{\mathcal{M}}$, where the operator $\{p|\bm{a}\}$ acts on the position $\bm{r}$ as $\{p|\bm{a}\} \bm{r} = p\bm{r} +\bm{a}$, while $\bm{t}_{\sigma}$ and $\bm{t}_{\theta}$ correspond to the zero or half translation operators. 
We note that any non-symmorphic generalizations of $I$, $\theta$, $\sigma_h$, and $C_2$ can be written in the above forms by an appropriate choice of the origin.
Using the generalizations provided above, the operators $\mathcal{\hat{T}}$, $\mathcal{\hat{M}}$, and $\mathcal{\hat{C}}_2$ include magnetic translation, glide, and screw transformations, respectively, and thus allow for the complete classification of symmetry-protected line nodes.

The characters of $P_{\bm{k}}$ can thus be obtained using Eq.~(\ref{eq:MB}), and are as listed in Table~\ref{tab:character}~\cite{suppl}, where $[\bm{t}_{\sigma}]_{\perp}$ and $[\bm{t}_{\theta}]_{\perp}$ correspond to translations perpendicular to the mirror plane. We find that the results can be classified into four different cases~\cite{Mickliz17PRL}: (a) $[\bm{t}_{\sigma}]_{\perp}= [\bm{t}_{\theta}]_{\perp}=0$; (b) $[\bm{t}_{\sigma}]_{\perp} \neq 0$ and $[\bm{t}_{\theta}]_{\perp}=0$; (c) $[\bm{t}_{\sigma}]_{\perp}=0$ and $[\bm{t}_{\theta}]_{\perp}\neq 0$; and (d) $[\bm{t}_{\sigma}]_{\perp}\neq 0$ and $[\bm{t}_{\theta}]_{\perp} \neq 0$. 
In the basal plane (BP) ($k_{\perp}=0$, where $k_{\perp}$ is the momentum normal to the mirror plane), the symmetry operators have the common expected characters in all cases, but on the Brillouin ZF ($k_{\perp}=\pi$), the characters in each case are different. Comparing these characters with those of the IRs in Table~\ref{tab:character}, we can thus obtain the IR decomposition of $P_{\bm{k}}$, which is summarized in Table~\ref{tab:group} and indicates the possible pairing symmetries of Cooper pairs that are consistent with the crystal structure. 
For instance, $A_g$ indicates the conventional $s$-wave pairing and is always possible in the $k_{\perp}=0$ plane. On the other hand, in cases (c) and (d), $A_g$ pairing is forbidden in the $k_{\perp} = \pi$ plane due to magnetic translation effects, indicating that in these cases, a line node appears on the ZF even for an $s$-wave SC~\cite{Nomoto17,Sumita17}.        

\begin{table}[tbp]
\centering
 \caption{Character table of $P_{\bm{k}}$ in the two possible mirror planes, $k_{\perp} =0$ and $k_{\perp} =\pi$, where $k_{\perp}$ is the momentum perpendicular to the mirror plane, $[\bm{t}_{\sigma}]_{\perp}$ and $[\bm{t}_{\theta}]_{\perp}$ represent translations normal to the mirror plane, and $\mathcal{\hat{E}}$ is the unit element of the crystal group symmetry. The bottom table shows the IRs and their group characters generated by ($\mathcal{\hat{E}}$, $\hat{\mathcal{C}}_2$, $\hat{\mathcal{P}}$, $\hat{\mathcal{M}}$), i.e., $C_{2h}$. \vspace*{2mm}} \label{tab:character}
\begin{tabular}{cc|ccccc} 
\hline \hline
&$P_{k_{\perp} =0}$ & $\mathcal{\hat{E}}$ & $\hat{\mathcal{C}}_2$ & $\hat{\mathcal{P}}$ & $\hat{\mathcal{M}}$ \\ \hline
&$^{\forall}\bm{t}_{\theta}, \bm{t}_{\sigma}$ & $4$ & $2 $ & $-2$ & $0$ \\
\hline \hline
&$P_{k_{\perp} =\pi}$ & $\mathcal{\hat{E}}$ & $\hat{\mathcal{C}}_2$ & $\hat{\mathcal{P}}$ & $\hat{\mathcal{M}}$ \\ \hline
(a)& $[\bm{t}_{\sigma}]_{\perp}= [\bm{t}_{\theta}]_{\perp}=0$ & $4$ & $2 $ & $-2$ & $0$ \\
(b)& $[\bm{t}_{\sigma}]_{\perp} \neq 0$, $[\bm{t}_{\theta}]_{\perp}=0$ & $4$ & $-2$ & $-2$ & $4$ \\
(c)& $[\bm{t}_{\sigma}]_{\perp}=0$, $[\bm{t}_{\theta}]_{\perp}\neq 0$ & $4$ & $2$ & $-2$ & $-4$ \\
(d) &$[\bm{t}_{\sigma}]_{\perp}\neq 0$, $[\bm{t}_{\theta}]_{\perp} \neq 0$ & $4$ & $-2$ & $-2$ & $0$ \\
\hline \hline
&IRs & $\mathcal{\hat{E}}$ & $\hat{\mathcal{C}}_2$ & $\hat{\mathcal{P}}$ & $\hat{\mathcal{M}}$ \\ \hline
&$A_g$ &$1$&$1$&$1$&$1$  \\
&$B_g$&$1$&$-1$&$1$&$-1$   \\
&$A_u$&$1$&$1$&$-1$&$-1$ \\
&$B_u$&$1$&$-1$&$-1$&$1$  \\
  \hline \hline
\end{tabular}
\end{table}

\begin{table}[tbp]
\centering
 \caption{IR decompositions of $P_{\bm{k}}$ in the mirror planes $k_{\perp} =0$ and $ k_{\perp}=\pi$.\vspace*{2mm}} \label{tab:group}
\begin{tabular}{c|ccc} 
\hline \hline
Cases & $P_{k_{\perp}=0}$ &  & $P_{k_{\perp}=\pi}$ \\ \hline
(a) & $A_g+2A_u+B_u$& $\to$ & $A_g+2A_u+B_u$ \\
(b) & $A_g+2A_u+B_u$& $\to$& $A_g+3B_u$ \\
(c) & $A_g+2A_u+B_u$& $\to$& $B_g+3A_u$   \\
(d) & $A_g+2A_u+B_u$& $\to$& $B_g+A_u+2B_u$  \\
  \hline \hline
\end{tabular}
\end{table}

{\it Topology of symmetry-protected line nodes}---Let us now discuss the symmetry-protected line nodes from the viewpoint of topology. Such topological properties are of importance in identifying the bulk--boundary correspondence, i.e., the line-node-induced MFBs. We can formulate the topology of line nodes using the Bogoliubov--de Gennes (BdG) Hamiltonian, $H_{\rm BdG} = \frac{1}{2} \sum_{\bm{k},\alpha,\beta} \Psi_{\bm{k},\alpha}^{\dagger} \tilde{H}(\bm{k})_{\alpha \beta}  \Psi_{\bm{k},\beta}$, where $\Psi^{T}_{\bm{k},\alpha} = (c_{\bm{k},\alpha},c_{-\bm{k},\alpha}^{\dagger})$ 
 and
\begin{align}
 \tilde{H}(\bm{k})_{\alpha \beta}  = \begin{pmatrix} H(\bm{k})_{\alpha \beta}-\mu \delta_{\alpha \beta}  & \Delta (\bm{k})_{\alpha \beta}  \\ \Delta(\bm{k})_{\alpha \beta} ^{\dagger} & -H(-\bm{k})_{\alpha \beta} ^{T} +\mu \delta_{\alpha \beta}\end{pmatrix}. \label{eq:BdGH}
\end{align}
Here, $H(\bm{k})$, $\Delta(\bm{k})$, and $\mu$ are the normal Hamiltonian, gap function, and chemical potential, respectively.  
We note that this Hamiltonian exhibits particle-hole (PH) symmetry since $C\tilde{H}(\bm{k})C^{\dagger} =-\tilde{H}(-\bm{k})$, where $C=\tau_x K$ is the anti-unitary operator, $\bm{\tau}$ is the Pauli matrix in Nambu space, and $K$ is the complex conjugate. We choose a periodic Bloch basis so that $\tilde{H}(\bm{k})=\tilde{H}(\bm{k}+\bm{G})$, where $\bm{G}$ a reciprocal lattice (RL) vector.

First, we consider how the symmetry operations affect the BdG Hamiltonian. The creation operator of an electron satisfies $\{p|\bm{a}_p\} c_{\bm{k},\alpha}^{\dagger} \{p|\bm{a}_p\}^{-1}= c_{p \bm{k},\beta}^{\dagger} [e^{-i p\bm{k} \cdot \bm{a}_p}U(p)_{\bm{k}} ]_{\beta \alpha}$, where $\{p|\bm{a}_p\} = \mathcal{\hat{P}}$, $\mathcal{\hat{M}}$, or $\mathcal{\hat{C}}_2$. If the BdG Hamiltonian is invariant with respect to the symmetry operations, Eq.~(\ref{eq:BdGH}) yields
$
 \tilde{U}(p)_{\bm{k}} \tilde{H}(\bm{k}) \tilde{U} (p)_{\bm{k}}^{\dagger}= \tilde{H}(p\bm{k}) 
$
with $\tilde{U}(p)_{\bm{k}} =\diag[U_{\bm{k}}(p), \eta_{p} U(p)_{-\bm{k}}^{\ast}]$.
Here, $\eta_p=\pm 1$, where the choice of sign is the same as the sign of the character of $p$ in the IR of $\Delta (\bm{k})$.
In addition, $\mathcal{\hat{T}}$ acts on the creation operators in a similar way to TR symmetry, and is thus accompanied by an additional momentum factor for $\bm{t}_{\theta} \neq 0$,  i.e., $ \mathcal{\hat{T}} c_{\bm{k},\alpha}^{\dagger} \mathcal{\hat{T}}^{-1}= c_{- \bm{k},\beta}^{\dagger} [e^{i\bm{k} \cdot \bm{t}_{\theta}}U(\mathcal{\theta})_{\bm{k}} ]_{\beta \alpha}$, which yields 
 $
\tilde{U}(\theta)_{\bm{k}} \tilde{H}(\bm{k})^{\ast} \tilde{U} (\theta)_{\bm{k}}^{\dagger}= \tilde{H}(-\bm{k}),
$
with  $\tilde{U}(\theta)_{\bm{k}} = \diag [U(\theta)_{\bm{k}}, U(\theta)_{-\bm{k}}^{\ast}]$.

Next, we clarify the relations between the symmetry operations. On the mirror plane, the PH operator satisfies
\begin{align}
 C\tilde{U}(p)_{\bm{k}} = \eta_p \tilde{U}(p)_{-\bm{k}} C, \label{eq:CP}
 \end{align}
 while $U(\sigma_h)_{\bm{k}}$, $U(I)_{\bm{k}}$, and $U(\theta)_{\bm{k}}$ satisfy 
\begin{subequations}
\label{eq:factor} \begin{align}
 &U(I)_{\bm{k}} U(\sigma_h)_{\bm{k}} = \eta_{I,\sigma_h}U(\sigma_h)_{-\bm{k}}U(I)_{\bm{k}}, \\
 &U(\theta)_{\bm{k}} U(\sigma_h)_{\bm{k}}^{\ast} = \eta_{\theta,\sigma_h}e^{i (\sigma_h \bm{k}-\bm{k}) \cdot \bm{t}_{\theta} } U(\sigma_h)_{-\bm{k}} U(\theta)_{\bm{k}},
\end{align}
\end{subequations}
where we have used $U(pp')_{\bm{k}}=\eta_{p,p'}U(p'p)_{\bm{k}}$ and the additional phase factor $ e^{i (\sigma_h \bm{k}-\bm{k}) \cdot \bm{t}_{\theta} }$ (called the factor system~\cite{Bradley72}) is only nontrivial on the ZF.
\begin{table}[tbp]
\centering
 \caption{ (Color online) Topology of the symmetry-protected line nodes, labeled by the symbol $M^{ p\; q}_{g(u)}$. The superscripts $p,q$ are defined by $p=e^{-i \bm{G}_{\sigma} \cdot \bm{t}_{\sigma}} \, \eta_{\sigma_h} \eta_{I,\sigma_h}$ and $q=e^{-i\bm{G}_{\sigma} \cdot \bm{t}_{\theta}} \, \eta_{\sigma_h} \eta_{\theta,\sigma_h}$, respectively, while the subscript $g(u)$ indicates whether the SC has even (g) or odd (u) parity. The bottom table shows the comparison between Table~\ref{tab:group} and the $0$D topological numbers, where $A_g$($B_u$) and $B_g$($A_u$) in the BP correspond to the $M^{++}_{g(u)}$ and $M^{--}_{g(u)}$ classes, respectively. The labels (i)--(iii) indicate the SC class of the MFBs.  
  \vspace*{2mm}} \label{tab:topo}
 
 \begin{tabular}{c|c|c|c|c|c|c|c|c} 
\hline \hline
 Topo. \# & $M^{++}_g$ & $M^{+-}_g$ & $M^{-+}_g$ &$M^{--}_g$ &$M^{++}_u$ &$M^{+-}_u$ & $M^{-+}_u$ & $M^{--}_u$ \\ \hline
 0D&0&$2 \mathbb{Z}$&0&$\mathbb{Z}_2$&0&$2 \mathbb{Z}$&$0$&$0$ \\
 1D&$2 \mathbb{Z}$&$2 \mathbb{Z}$&$2 \mathbb{Z}$&$2 \mathbb{Z}$&$0$&$0$&$0$&$0$ \\
  \hline \hline
\end{tabular}

 \vspace{2mm}
 
\begin{tabular}{c|ccc|ccc|ccc|ccc} 
\hline \hline
 &  \multicolumn{3}{|c|}{$A_g$}  & \multicolumn{3}{c}{$B_g$}  &  \multicolumn{3}{|c|}{ $A_u$} &  \multicolumn{3}{c}{$B_u$} \\ \hline
Cases  &\ \ BP & & ZF& \ \  BP & & ZF & \ \ BP & & ZF& \ \  BP & & ZF \\ \hline
(a) & \ \ $0$  & $\to$ & $0$ &  \textcolor{blue}{(ii) $\mathbb{Z}_2$} & \textcolor{blue}{$\to$} &  \textcolor{blue}{$\mathbb{Z}_2$}&\ \ $0$  & $\to$ & $0$ &\ \ $0$ & $\to$ & $0$ \\
(b)  &\ \ $0$ & $\to$ &$0$ &  \textcolor{blue}{(ii) $\mathbb{Z}_2$}& \textcolor{blue}{$\to$} & \textcolor{blue}{$2\mathbb{Z}$} & \textcolor{green}{(i) $0$} & \textcolor{green}{$\to$} &\textcolor{green}{$2\mathbb{Z}$} & \ \ $0$& $\to$ & $0$ \\
(c) & \textcolor{red}{(iii) $0$} & \textcolor{red}{$\to$} & \textcolor{red}{$2\mathbb{Z}$} &  \textcolor{red}{(iii) $\mathbb{Z}_2$}& \textcolor{red}{$\to$} & \textcolor{red}{$0$}  &\ \ $0$ & $\to$ & $0$ & \textcolor{green}{(i) $0$}& \textcolor{green}{$\to$} & \textcolor{green}{$2\mathbb{Z}$}\\
(d) & \textcolor{red}{(iii) $0$} & \textcolor{red}{$\to$} &\textcolor{red}{$\mathbb{Z}_2$} &  \textcolor{red}{(iii) $\mathbb{Z}_2$}& \textcolor{red}{$\to$} & \textcolor{red}{$0$}  &\ \ $0$ & $\to$ &$0$ &\ \ $0$& $\to$ & $0$   \\
  \hline \hline
\end{tabular}
\end{table}
We assume that the line node lies on the mirror plane and consider a set of symmetry operations $\mathcal{S}$ that keep the position of the line node invariant, i.e., $\mathcal{S}:\bm{k} \to \bm{k} + \bm{G}$. The symmetry operations $\mathcal{S}$ then consist of the PH-like operator $\mathcal{C}_{\bm{k}} \equiv C \tilde{U}(I)_{\bm{k}}$ and the TR-like operator $\mathcal{T}_{\bm{k}} \equiv \tilde{U}(I)_{-\bm{k}}\tilde{U}(\theta)_{\bm{k}}K$, which are obtained from combinations of the PH and TR operators with inversion operators, and the chiral operator $\Gamma_{\bm{k}} \equiv iC\tilde{U}(\theta)_{\bm{k}}K$. Here, $\eta_I \mathcal{C}_{\bm{k}}^2 = -\mathcal{T}_{\bm{k}}^2 = \Gamma_{\bm{k}}^2 =1$. Moreover, the MR operation $\tilde{U}(\sigma)_{\bm{k}}$ is also included in $\mathcal{S}$ since $\sigma_h: \bm{k} \to \bm{k} + \bm{G}_{\sigma}$, where $\bm{G}_{\sigma}$ is the RL vector  normal to the mirror plane. Since $\tilde{U}(\sigma)_{\bm{k}}^2 = -e^{-i \bm{G}_{\sigma} \cdot \bm{t}_{\sigma}}$, the sign of the squared operator may also change at the ZF. We conveniently fix the sign by defining $\mathcal{M}_{\bm{k}} \equiv e^{i \bm{G}_{\sigma} \cdot \bm{t}_{\sigma}/2} \tilde{U}(\sigma)_{\bm{k}}$ which satisfies $\mathcal{M}_{\bm{k}}^2=-1$.
Using Eqs.~(\ref{eq:CP}) and (\ref{eq:factor}), we then obtain the commutation relations:
\begin{subequations}
\label{eq:CP-CT} \begin{align}
 &\mathcal{C}_{\bm{k}} \mathcal{M}_{\bm{k}} =e^{-i \bm{G}_{\sigma} \cdot \bm{t}_{\sigma}} \, \eta_{\sigma_h} \eta_{I,\sigma_h} \mathcal{M}_{\bm{k}} \mathcal{C}_{\bm{k}}, \label{eq:CP-M}\\
 &\Gamma_{\bm{k}} \mathcal{M}_{\bm{k}} =e^{-i\bm{G}_{\sigma} \cdot \bm{t}_{\theta}} \, \eta_{\sigma_h} \eta_{\theta,\sigma_h} \mathcal{M}_{\bm{k}} \Gamma_{\bm{k}}, \label{eq:CT-M}
\end{align}
\end{subequations}
while the commutation relation between $\mathcal{T}_{\bm{k}}$ and $\mathcal{M}_{\bm{k}}$ can be determined from Eq.~(\ref{eq:CP-CT}). Since $\bm{G}_{\sigma} \cdot \bm{t}_{p} = 2\pi [\bm{t}_p]_{\perp}$, the right-hand side of Eq.~(\ref{eq:CP-CT}) may change sign in the BP and the ZF, depending on the action of $[\bm{t}_{\sigma}]_{\perp}$ and $[\bm{t}_{\theta}]_{\perp}$. 
\begin{figure}[tbp]
\centering
 \includegraphics[width=8cm]{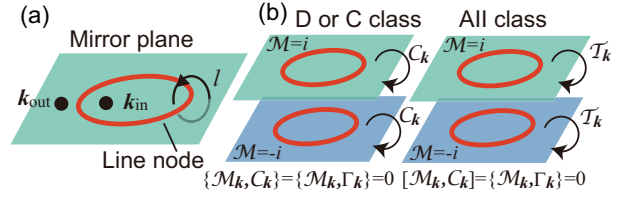}
 \caption{(Color online) (a) Schematic illustration of the 0D and 1D topological numbers. (b) Topological classes within the mirror sectors.}\label{fig:topo}
\end{figure}

1D and 0D topological numbers may exist and can be used to specify the line nodes. The former is defined on a loop encircling a line node irrespective of the MR symmetry, whereas the latter is defined at a $\bm{k}$ point on the mirror plane (see Fig.~\ref{fig:topo}(a)).  

First, let us consider the 1D topological number. Since $\mathcal{C}_{\bm{k}}^2= \eta_I$, even-parity SCs ($\eta_I=1$) and odd-parity SCs ($\eta_{I} = -1$) belong to different topological classes, and only even-parity SCs can support a non-zero 1D topological number~\cite{Kobayashi14,Zhao16,Bzdusek17}. In particular, the 1D topological number for even-parity SCs can be defined on a loop $l$ in terms of the chiral operator:
\begin{align}
 W_l=\frac{i}{4 \pi} \oint_l d \bm{k} \cdot \trace \left[ \Gamma_{\bm{k}} \tilde{H}(\bm{k})^{-1} \bm{\partial_{k}} \tilde{H}(\bm{k})\right]. \label{eq:1dw}
\end{align}

Next, we consider the $0$D topological number, for which we refer to the topological periodic table presented in~\cite{Schnyder08,Schnyder09,Kitaev09,Ryu10} and regarding $\mathcal{T}_{\bm{k}}$, $\mathcal{C}_{\bm{k}}$, and $\Gamma_{\bm{k}}$ in terms of the Altland--Zirnbauer symmetry~\cite{Zirnbauer96,Altland97}.  In terms of this classification, the BdG Hamiltonian at a point belongs to the DIII class for even-parity SCs and the CII class for odd-parity SCs, neither of which have a non-zero topological number. 
Thus, the MR symmetry is essential to the existence of a nontrivial topological number. We now show the existence of a $0$D topological number using the MR symmetry explicitly. In the mirror plane, the MR operator commutes with the BdG Hamiltonian and so the BdG Hamiltonian splits into MR sectors: $\tilde{H} \to \tilde{h}_{\lambda} \oplus \tilde{h}_{-\lambda}$, where $\lambda$ is an eigenvalue of $\mathcal{M}_{\bm{k}}$. Then, the symmetry operation $\Gamma_{\bm{k}}$ ($\mathcal{C}_{\bm{k}}$) exists within the MR sectors if it (anti-)commutes with $\mathcal{M}_{\bm{k}}$. The $0$D topological number is thus defined within the MR sectors, and moreover, when a chiral symmetry is present in the MR sectors, i.e., $[\Gamma_{\bm{k}},\mathcal{M}_{\bm{k}}]=0$, all cases are topologically trivial. Thus, we require $\{\Gamma_{\bm{k}},\mathcal{M}_{\bm{k}}\}=0$. In this case, then as shown in Fig.~\ref{fig:topo}(b), the MR sector belongs to one of the D, C, or AII classes, whose 0D topological numbers are $\mathbb{Z}_2$, $0$, and $2 \mathbb{Z}$, respectively. The $2\mathbb{Z}$ and $\mathbb{Z}_2$ numbers of a nodal loop (i.e., a loop-shaped line node) in the MR sector $h_{\lambda}$ can then be defined as
\begin{align}
 &\tilde{\mathcal{N}}_{\lambda} = \tilde{n}(\bm{k}_{\rm out})_{\lambda} - \tilde{n}(\bm{k}_{\rm in})_{\lambda}, \label{eq:mirrortopo} \\
  &(-1)^{\tilde{\nu}_{\lambda}} = \sgn \left[\frac{\pf \{ \tilde{h}_{\lambda} (\bm{k}_{\rm out})L_{\bm{k}_{\rm out},  \lambda}\}}{\pf \{\tilde{h}_{\lambda} (\bm{k}_{\rm in})L_{\bm{k}_{\rm in}, \lambda}\}}\right], \label{eq:mirrorZ2}
\end{align}
respectively, where $\tilde{n} (\bm{k})_{\lambda}$ is the number of occupied states with momentum $\bm{k}$, $\mathcal{C}_{\bm{k}} = (L_{\bm{k},\lambda} \oplus L_{\bm{k},-\lambda}) K$, and $\bm{k}_{\rm in (out)}$ is the momentum inside (outside) of the nodal loop. In the weak coupling limit, i.e., $\Delta(\bm{k}) \to 0$, Eqs.~(\ref{eq:mirrortopo}) and (\ref{eq:mirrorZ2}) are directly linked to the Fermi surface topology. If we define $\mathcal{N}_{\lambda} = n(\bm{k}_{\rm out})_{\lambda} - n(\bm{k}_{\rm in})_{\lambda}$ in the normal Hamiltonian to be the topological number of the Fermi surface in the mirror sector with eigenvalue $\lambda$, Eqs.~(\ref{eq:mirrortopo}) and (\ref{eq:mirrorZ2}) are reduced to   
$\tilde{\mathcal{N}}_{\lambda} = 2\mathcal{N}_{\lambda}$ and $(-1)^{\tilde{\nu}_{\lambda}} = (-1)^{\mathcal{N}_{\lambda}}$~\cite{suppl}, which implies that a nodal loop is only topologically stable if $\mathcal{N}_{\lambda} \neq 0$. We have summarized the possible topological numbers in Table~\ref{tab:topo}, where the symbol $M^{ p\; q}_{g(u)}$ encodes the signs of $p=e^{-i \bm{G}_{\sigma} \cdot \bm{t}_{\sigma}} \, \eta_{\sigma_h} \eta_{I,\sigma_h}$ and $q=e^{-i\bm{G}_{\sigma} \cdot \bm{t}_{\theta}} \, \eta_{\sigma_h} \eta_{\theta,\sigma_h}$, and the parity of the gap function, i.e., $g(u)$ indicates the even(odd)-parity SCs~\cite{note1}.

Thus, we are now in a position to elucidate the relationship between the group theoretical and topological classifications outlined above. For convenience, we assume that $\eta_{I,\sigma_h}=\eta_{\theta,\sigma_h}=1$ so that almost all SCs qualify for our description. The IRs $A_g$, $A_u$, $B_g$, and $B_u$ then correspond to the topological classifications labeled by $M^{++}_g$, $M^{--}_u$, $M^{--}_g$, and $M^{++}_u$ in the BP. Moreover, the MR symmetry classes in the ZF depend on $[\bm{t}_{\sigma}]_{\perp}$ and $[\bm{t}_{\theta}]_{\perp}$ due to Eq.~(\ref{eq:CP-CT}). In Table~\ref{tab:topo}, we have summarized the correspondence between the $0$D topological numbers and the four cases, (a)--(d). In comparison to Tables~\ref{tab:group} and \ref{tab:topo}, we find a one-to-one correspondence, in which the absence of IRs coincides with the presence of the 0D topological numbers.  

{\it Possible MFBs}---Finally, we consider the connection between symmetry-protected line nodes and MFBs, which are characterized by 0D and 1D topological numbers, respectively. As was discussed above, a 1D topological number exists when $\eta_I=1$, so MFBs only appear in the case of even-parity SCs.
In such even-parity SCs, the topological numbers are intrinsically related to one another and satisfy~\cite{suppl}
\begin{align}
 |\tilde{\mathcal{N}}_{\lambda} | = |W_l|, \ \ (-1)^{\tilde{\nu}_{\lambda}} = (-1)^{\frac{W_l}{2}}, \label{eq:rel0-1D}
\end{align}
which implies that the 1D topological number is always associated with the existence of symmetry-protected line nodes. We note also that the 1D topological number survives even when the MR symmetry is absent, which reflects the strong stability of the line nodes. 

Using Table~\ref{tab:topo} and Eq.~(\ref{eq:rel0-1D}), we can identify three classes with respect to the stability of the line nodes. A line node may be protected by: (i) the $0$D topological number in odd-parity SCs with TR or a magnetic translation symmetry, or alternatively, by both the $0$D and $1$D topological numbers in even parity SCs with (ii) TR or (iii) a magnetic translation symmetry. We immediately find that there is no MFB in Class (i) SCs since there is no 1D topological number corresponding to a MFB in odd-parity SCs. In order to demonstrate the existence of MFBs in Class (ii) and (iii) SCs, 
consider a system with an open boundary, e.g., the $x_i=0$ plane. Then, an MFB appears on the $x_i=0$ plane if the 1D topological number (\ref{eq:1dw}) defined on the loop $l(k_j,k_l) = \{(k_i,k_j,k_l)| -\pi \le k_i \le \pi$\}
is nonzero~\cite{Ryu02,Sato11}, where $(k_i,k_j,k_l)$ are perpendicular to each other, and $l(k_j,k_l)$ does not intersect with the line node.
For SCs satisfying the conditions of Class (ii), the operator $\hat{\mathcal{T}}$ corresponds to a pure TRS, so $l(k_j,k_l)$ can be defined for arbitrary surface direction. Thus, an MFB appears on the SC's surface in analogy with high-$T_{\rm c}$ cuprate SCs~\cite{Tanaka02,Tanaka12}. On the other hand, $\hat{\mathcal{T}}$ in Class (iii) corresponds to a magnetic translation, so $l(k_j,k_l)$ needs to be compatible with a translation vector $\bm{t}_{\theta}$, i.e., an MFB only arises when one satisfies
\begin{align}
 \bm{t}_{\theta} \cdot \hat{\bm{e}}_i=0 \ \ \text{for Class (iii) SCs}, \label{eq:magMFB}
\end{align}
where $\hat{\bm{e}}_i$ is a unit vector normal to the surface. Note that the behavior we have outlined here is similar to that of surface states in aniferromagnetic topological insulators~\cite{Mong10,Turner12,Essin12,Miyakoshi13,Yoshida13,Baireuther14,Zhang15,Frederic17}.  
Thus, a limitation on possible MFBs in Class (iii) SCs appears in contrast to MFBs in Class (ii) and non-centrosymmetric SCs~\cite{Brydon11,Yada11,Schnyder12}. 
In particular, when $\bm{t}_{\theta}$ is perpendicular to the mirror plane, MFBs do not exist on any surface because no surface direction simultaneously satisfies the constraints arising from Eq.~(\ref{eq:magMFB}) and the MR symmetry. Thus, a distortion or interaction that breaks the MR symmetry is necessary to reveal the hidden MFBs.

{\it Application to UPd$_2$Al$_{3}$}---In order to verify the existence of a MFB in Class (iii) SCs, let us consider a minimal model of the antiferromagnetic SC UPd$_2$Al$_{3}$~\cite{Nomoto17}. The antiferromagnetic phase of this material is specified by the magnetic space group $P_b 2_1/m$, and the two U atoms are situated at the $(0,0,0)$ and $(0,0,\frac{1}{2})$ positions in the magnetic unit cell. 
\begin{figure}[tbp]
\centering
 \includegraphics[width=8cm]{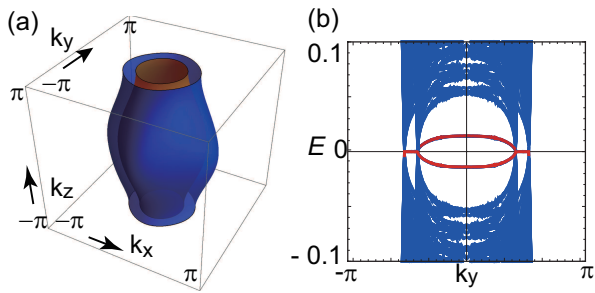}
 \caption{(Color online) (a) Fermi surface of the model (\ref{eq:UPd2A3}) with parameters $(t_{xy}, t_z, t_z', t_{xz}, \delta_M, \Delta_0, \mu)=(1,0.4,0.1,0.1,0.4,0.2,-2)$. (b) Energy spectrum in the $(100)$ plane, where the thick red line indicates the surface state.}\label{fig:UPd2Al3}
\end{figure}
Taking a single orbital at each U site into account, the tight-binding model can then be given by
\begin{align}
 H(\bm{k}) = \epsilon(\bm{k})_1  + \epsilon (\bm{k})_2 \sigma_x(k_z) + \delta_M \sigma_z s_x , \label{eq:UPd2A3}
\end{align}
where $\epsilon (\bm{k})_1 = -2 t_{xy}(\cos k_x + \cos k_y) - 2 t_z' \cos k_z $, $\epsilon (\bm{k})_2 = -2 t_z \cos \frac{k_z}{2}$, and $\sigma_x(k_z) \equiv \cos \frac{k_z}{2} \sigma_x + \sin \frac{k_z}{2} \sigma_y$. The constants $t_{xy}$, $t_z'$, and $t_z$ are hopping parameters and $\delta_M$ represents the molecular field in terms of the magnetic moment. For simplicity, the gap function is assumed to be given by the $s$-wave spin-singlet pairing, $\Delta_0 i s_y$, and the symmetry operators are then given by $\mathcal{C}_{\bm{k}} = e^{i\frac{k_z}{2}}(\cos \frac{k_z}{2} \sigma_0 - i \sin \frac{k_z}{2} \sigma_z)\tau_x K$, $\mathcal{T}_{\bm{k}} = e^{i \frac{k_z}{2}} i \sigma_x s_y K $, $\Gamma_{\bm{k}} = - \sigma_x (k_z) s_y \tau_x$, and $\mathcal{M}_{\bm{k}} = i\sigma_x s_z \tau_z$, where $\bm{\sigma}$ and $\bm{s}$ are the Pauli matrices in the sublattice and spin spaces, respectively. This model is known to host two nodal loops in the $k_z = \pi$ plane~\cite{Nomoto17}. Since the MR symmetry class is $M^{--}_g$, the nodal loops are protected by both 0D and 1D topological numbers. 
In fact, using Eq.~(\ref{eq:rel0-1D}), we find $\nu_{\lambda}=1$ and $|W_l| = 2$ for each nodal loop and the resulting nontrivial Fermi surface topology is shown in Fig.~\ref{fig:UPd2Al3}(a)~\cite{suppl}. Notably, MFBs are absent in all surface planes because $\bm{t}_{\theta} = \frac{1}{2} \hat{\bm{e}}_z$ is perpendicular to the mirror plane.
 
However, when we add an MR-breaking distortion term, e.g., $2 t_{xz} \sin k_x \sin \frac{k_z}{2} \sigma_x(k_z) $ to the Hamiltonian, the nodal loop escapes from the mirror plane and a gap opens on the $k_z=\pi$ plane. 
In this case, the surface state can be numerically determined and the MFBs are revealed by the MR-breaking distortion, as shown in Fig.~\ref{fig:UPd2Al3}(b).

{\it Concluding remarks}---We have established the relationship between symmetry-protected line nodes and MFBs using a topological argument.
As revealed in Tables~\ref{tab:group} and~\ref{tab:topo}, the 0D topological number not only reflects the group theoretical results, but also relates to the 1D topological number that ensures the existence of MFBs.     
By analyzing the relationship between the two topological numbers, we can categorize the symmetry-protected line nodes into three distinct classes, which we label Classes (i)--(iii), as summarized in Table~\ref{tab:topo}. Each class is distinguished by the MFBs and the symmetry-protected line nodes may be distinguished through surface sensitive experiments such as tunneling conductance measurements.

The authors are grateful to A. Yamakage for useful discussions.
This work was supported by the Grants-in-Aid for Scientific Research on Innovative Areas ``J-Physics'' (Grant No. JP15H05884) and ``Topological Material Science'' (Grant Nos. JP15H05855 and JP16H00991) from JSPS of Japan, and by JSPS KAKENHI Grant Nos. 15K05164, JP15H05745, JP16H06861, JP17H02922, and JP17J09908. S.K. was supported by the Building of Consortia for the Development of Human Resources in Science and Technology.
\bibliography{line}

\begin{thebibliography}{69}
\bibitem [{not()}]{note1}%
  \BibitemOpen
  \href@noop {} {}\bibinfo {note} {The same topological numbers are obtained by
  solving the Clifford algebra extension problem~\cite{suppl}.}\BibitemShut
  {Stop}%
\end{thebibliography}
\end{document}